\documentclass[prodmode,acmcsur]{acmsmall}

\usepackage[ruled]{algorithm2e}
\renewcommand{\algorithmcfname}{ALGORITHM}
\SetAlFnt{\small}
\SetAlCapFnt{\small}
\SetAlCapNameFnt{\small}
\SetAlCapHSkip{0pt}
\IncMargin{-\parindent}

\usepackage{color}
\usepackage{graphicx}
\usepackage{subfigure}
\usepackage{url}
\usepackage{caption}

\makeatletter
\def\ps@headings{%
\def\@oddhead{\mbox{}\scriptsize\rightmark \hfil \thepage}%
\def\@evenhead{\scriptsize\thepage \hfil \leftmark\mbox{}}%
\def\@oddfoot{}%
\def\@evenfoot{}}
\makeatother
\newcommand{\ignore}[1]{}

\newcommand{\red}[1]{\textcolor{black}{#1}}

\begin{document}

\markboth{V. Pejovic and M. Musolesi}{Anticipatory Mobile Computing}

\title{Anticipatory Mobile Computing: A Survey of the State of the Art and Research Challenges}

\author{VELJKO PEJOVIC
\affil{School of Computer Science, University of Birmingham, UK}
MIRCO MUSOLESI
\affil{School of Computer Science, University of Birmingham, UK}}

\begin{abstract}
Today's mobile phones are far from mere communication devices they were ten years ago. Equipped with sophisticated sensors and advanced computing hardware, phones can be used to infer users' location, activity, social setting and more. As devices become increasingly intelligent, their capabilities evolve beyond inferring context to predicting it, and then reasoning and acting upon the predicted context. \red{ This article provides an overview of the current state of the art in mobile sensing and context prediction paving the way for full-fledged anticipatory mobile computing.} We present a survey of phenomena that mobile phones can infer and predict, and offer a description of machine learning techniques used for such predictions. We then discuss proactive decision making and decision delivery via the user-device feedback loop. Finally, we discuss the challenges and opportunities of anticipatory mobile computing.\end{abstract}

\category{A.1}{Introductory and Survey}{}
\category{H.3.4}{Systems and Software}{Distributed systems}
\category{H.1.2}{User/Machine Systems}{}
\category{I.2.6}{Learning}{}
\category{J.4}{Social and Behavioral Sciences}{}

\terms{Design, Human Factors, Performance}

\keywords{Anticipatory Computing, Mobile Sensing, Context-aware Systems}

\acmformat{Pejovic, V., Musolesi, M. 2014. Anticipatory Mobile Computing: A Survey of the State of the Art and Research Challenges}

\begin{bottomstuff}
This work was supported through the EPSRC grant ``UBhave: ubiquitous and social computing for positive behaviour change" (EP/I032673/1).

Authors' address: V. Pejovic {and} M. Musolesi, School of Computer Science,
University of Birmingham, Edgbaston B15 2TT Birmingham, United Kingdom.
\end{bottomstuff}
\maketitle

\section{Introduction}
\label{sec:introduction}

The ability to communicate on the move has revolutionised the lifestyle of millions of individuals: it has changed the way we work, organise our daily schedules, develop and maintain social ties, enjoy our free time, and handle emergencies. In the past decade mobile phones have reached every part of the world, and 86\% of the world's population had a cellular subscription in year 2012 \cite{ITU2012}. When smartphones replaced feature phones another mobile revolution happened. Nowadays, phones serve for travel planning, staying in touch with online social network contacts, online shopping and numerous other purposes. Today's smartphones with multi-core CPUs and gigabytes of memory are capable of processing tasks that yesterday's desktop computers struggled with. However, unlike desktop computers, smartphones are small mobile devices. Consequently, phones became a part of everyday life and remain continuously present and used at all times. In addition, modern-day smartphones host a variety of sophisticated sensors: a phone can sense its orientation, acceleration, location, and can record audio and video. As such, a smartphone is not just a mobile computer -- it is a perceptive device capable of extending human senses~\cite{Lane2010}. Finally, these devices are connected to the Internet and, therefore, they can share the collected data and exploit the resources offered by cloud services.

Despite the recent phenomenal progress, the area of mobile personal devices promises further advances as sensing and processing capabilities of mobile phones grow. \red{In this survey we discuss the emergence of \textit{anticipatory mobile computing}, a field that harnesses mobile sensing and machine learning for intelligent reasoning based on the prediction of future events.} We build this new paradigm upon the theoretical postulates of anticipatory systems -- computing systems that base their actions on a predictive model of themselves and their environment. Smartphones are potentially a revolutionary platform for anticipatory systems as they bridge the gap between the device, the environment and the user. First, they fulfil the necessary prerequisites for successful anticipatory reasoning: they are equipped with numerous sensors and can infer and monitor the context, while powerful processing hardware allows them to run machine learning algorithms and develop sophisticated models of the future. Second, phones are very closely integrated with everyday life of individuals \cite{Katz1998}. Thus, models developed on mobile phones can be very personal, timely and relevant for the user. In addition, interaction with the environment, which is crucial for the realisation of anticipatory decisions, is naturally supported due to the user's reliance on smartphone-provided information.

Anticipatory mobile computing is inherently interdisciplinary. Mobile sensing, human-computer interaction (HCI), machine learning, and context prediction are major research fields related to anticipatory mobile computing. Each of these areas is thoroughly covered in the existing survey literature, such as \cite{Butz2003,Chen2000,Lane2010,Burbey2012,Lanzi2008}, so we concentrate on an orthogonal goal and examine the role of each of the stages in the process of designing anticipatory mobile systems. Still, when necessary we systematically present developments in these subfields in order to provide a practitioner with an overview of possible implementation options. Overall, our goal is not only to give a thorough overview of the state of the art, but also to sketch practical guidelines for building anticipatory mobile systems.

We note that anticipatory computing is an often misused term, especially when it comes to describing the recent wave of context-aware and predictive applications for mobile devices. In the first part of this survey (Section~\ref{sec:anticipatory_glance}) we embrace and examine a well established definition of anticipatory computing~\cite{Rosen1985} stating that only applications that rely on past, present and anticipated future in order to make judicious actionable decisions can be considered anticipatory applications. We then argue that the smartphone is a true enabler of anticipatory computing (Section~\ref{sec:anticipatory_computing}). One of the smartphone's main affordances is the ability to sense an abundance of information about the environment. Therefore, we dedicate a part of the survey to mobile sensing and context inference (Section~\ref{sec:sensing_modelling}) from the point of view of anticipatory systems. These processes aim to reconstruct key characteristics of the user behaviour and the environment from sensed signals~\cite{Coutaz2005}. For reliable reconstruction, in each of the domains, whether it is speech analysis, movement tracking, object recognition or any other domain, we need to identify features of raw signals that are useful for inferring higher-level concepts and characteristics. We describe how information flows from the physical environment through phone's sensors, and gets processed by machine learning algorithms so that high-level information is extracted. The ability to infer the context in which it is operating makes a phone more than a communication device -- it becomes a \textit{sense}~\cite{Campbell2012}. Although the area of mobile sensing remains far from being fully explored, recent research is increasingly focused on providing \textit{cognitive} capabilities to mobile phones. This allows the phone to be trained to \textit{predict} future events from current and past sensor data. The ability to predict users' location, social encounters or health hazards pushes the smartphone further to an irreplaceable source of personalised information. While inferring usage context on the smartphone remains difficult due to the sheer amount and variable quality of highly user-specific data, predicting the future context is even more difficult. Context prediction is tied with problems such as identifying and gathering data relevant for prediction, and determining prediction reliability, prediction horizon and possible outcomes. The later part of this survey provides an overview of the existing work in context prediction with smartphones (Section~\ref{sec:context_prediction}). 
Finally, in the true sense of anticipatory computing, predictions made with the help of data gathered through mobile sensing can be used as a basis for intelligent decision making. In Section~\ref{sec:closing_loop} we investigate anticipatory mobile computing systems, that is, systems that rely on past, present and anticipated future in order to make judicious decisions about their actions. Ideas about computing devices that can autonomously adapt their performance over time is not new~\cite{Kephart2003}. With smartphones we are for the first time able to realise personalised anticipatory computing on a large scale (Section~\ref{sec:large_scale}). However, this also means that novel issues arise; these challenges for anticipatory mobile computing are examined in Section~\ref{sec:challenges}. 

\section{Overview of Anticipatory Systems: Definitions and Applications}

\label{sec:anticipatory_glance}

In this section we discuss a possible definition of anticipatory systems and the application of this class of systems to three different domains.

\subsection{Defining Anticipatory Systems} 

An anticipatory system is defined by Rosen as: 
\begin{quotation}
\em{``A system containing a predictive model of itself and/or its environment, which allows it to change state at an instant in accord with the model's predictions pertaining to a later instant"}~\cite{Rosen1985}. 
\end{quotation}
This definition hints that an anticipatory device needs to be capable of obtaining a realistic picture of its state and the surrounding environment, i.e., the context in which the user and the device are. Equipped with an array of sensors and powerful processing hardware that can support sophisticated machine learning algorithms, smartphones can build predictive models of the context. Anticipatory actions that impact the future state are then based on the predictions of the future state of the context. Tightly integrated with users' lifestyle, a phone can learn personalised patterns of their behaviour, and with the help of rich user interface it can communicate anticipatory actions to the users.

\subsection{Anticipatory Mobile Computing Applications} 

To illustrate the potential of smartphone-based anticipatory mobile computing here we present three example applications.
\subsubsection{Personal Assistant Technology} A mobile phone has access to a wealth of personal information, including Web browsing history, calendar events, and online social network contacts. Application developers can tap into this data and design applications that predict users' intentions and display momentary relevant content. MindMeld is one such application that enhances online video conferencing with information that the users are likely to find relevant in near future~\cite{MindMeld}. For this purpose MindMeld harnesses real-time speech analysis, machine learning and WWW harvesting. Google Now takes a more general approach and aims to provide a mobile phone user with any information or functionality she may need, without the user explicitly asking for it~\cite{GoogleNow}. If augmented with a model that anticipates environment's reaction to user's actions these predictive applications could become intelligent anticipatory personal assistants that perform autonomously for user's benefit. Such an application could, for example, foresee an encounter with one's business partners and prepare documents for a successful impromptu meeting. 

\subsubsection{Healthcare} Mobile sensing has been proposed as a means of providing in situ diagnosis~\cite{Gruenerbl2014}. In addition, mobile phones are increasingly being used to deliver personalised therapies~\cite{Klasnja2009,Morris2010,Yardley2013}. Currently these therapies tend to be pre-loaded on users phones and react according to the sensed context. Anticipatory computing can be used to build and develop a model of human behaviour and devise therapies automatically, aiming to lead the user towards a certain well-being goal. For example, through a built-in accelerometer the phone can sense user's level of physical activity and, by means of a Bluetooth sensor and calling behaviour, it can sense user's sociability. Then, the phone can infer the well-being state and predict if the user is in risk of major depression. Finally, it can adjust the therapy on-the-fly, for example by sending a link to two discounted theatre tickets, incentivising the participant to go out and socialise. Such a self-contained application that anticipates changes in user's health and behaviour allows scalability and personalisation unimaginable in the traditional physician-patient world.

\subsubsection{Smart Cities} The ratio of urban population experiences a steady growth and nowadays more than a half of approximately seven billion people living on Earth reside in urban areas~\cite{WHO2010}. Issues such as traffic, pollution and crime plague modern cities. Participatory mobile sensing, where citizens are actively involved in data collection, as well as opportunistic mobile sensing, where users simply volunteer to host an autonomous application on their devices, are already being employed for tackling urban problems\footnote{For an overview of urban sensing and human-centric sensing research, we refer the reader to~\cite{Campbell2008}, \cite{Srivastava2012} and \cite{EvansCowley2010}.}. For instance, MIT's CarTel project uses mobile sensing for traffic mitigation, road surface monitoring and hazard detection \cite{Hull2006}; ParkNet system collects parking space occupancy information through distributed sensing from passing-by vehicles \cite{Mathur2010}; Dutta et al. demonstrate a participatory sensing architecture for monitoring air quality \cite{Dutta2009}. An anticipatory system that makes autonomous decisions and reasons about their consequences can push such projects further. Thus, we envisage a smart navigation system that predicts traffic jams and directs drivers in order to alleviate road congestion and balance pollution levels across the city. 

%
%
%
%
%
%
%

\begin{figure}[t]
    \centering
    \includegraphics[width=0.99\textwidth]{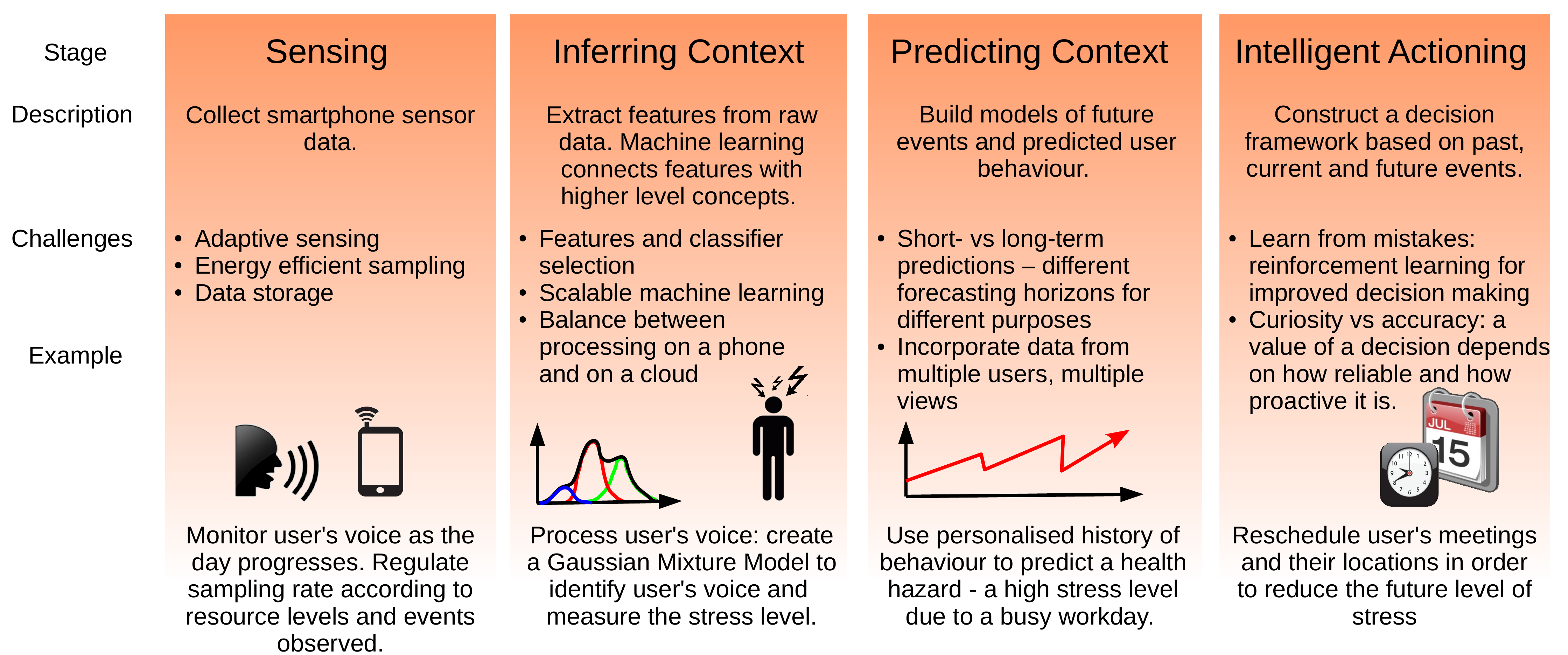}
    \caption{Key stages in the anticipatory mobile computing depicted on an example stress management application: collecting sensor data, processing it in order to infer context, predicting future events and using past, present and future to make intelligent autonomous decisions.}
    \label{fig:overview}
\end{figure}

To demonstrate the challenges of bringing the applications such as the above to life, and to show possible solutions, in Figure~\ref{fig:overview} we sketch a fictional application, inspired by StressSense~\cite{Lu2012}. This proactive stress management application unobtrusively monitors social signals~\cite{Vinciarelli2012}, such as the voice of a busy user, infers current stress levels from voice features, predicts future ones based on the user's calendar and then intelligently reschedules meetings so that the anticipated stress level is within healthy boundaries. We dissect the application with respect to the implementation stages: context sensing and inference, context prediction, and intelligent actioning. In the rest of the paper we will discuss main developments and key challenges in each of the stages.

\section{Designing and Implementing Anticipatory Mobile Systems}
\label{sec:anticipatory_computing}

In this section we discuss the concept of anticipatory behaviour and we present a general architecture of anticipatory computing systems.

\subsection{Anticipatory Behaviour and Anticipatory Computing Systems}

Anticipatory behaviour is defined by Butz, Sigaud and Gerard as: 
\begin{quotation}
\em{``a process or behaviour, that does not only depend on past and present but also on predictions, expectations, or beliefs about future''}~\cite{Butz2003}. 
\end{quotation}
This behaviour is natural, in the sense that it is deeply integrated with intelligence, and biological systems often base decisions for their actions on predictions~\cite{Rosen1985}. An animal increases its chance of survival by predicting a dangerous situation, a tennis player hits a ball on time by predicting its trajectory, and the prediction of rain helps us carry an umbrella and stay dry. Anticipatory behaviour has been confirmed in experimental psychology~\cite{Tolman1932}, while neuropsychology has provided further insights about brain mechanics related to anticipation~\cite{Gallese1998}.


Are computing devices capable of implementing brain functions and mimicking the mind when it comes to anticipation? A positive answer would lead to the realisation of anticipatory behaviour in a computing system and open up tremendous opportunities for exploiting such capabilities in applications ranging from personal assistants to healthcare and robotics. The past three decades saw substantial efforts in the area of anticipatory computing, with the goal of bringing anticipatory computing systems, as defined by Rosen, to life. \red{During this time milestones such as the formalisation of anticipatory computing system architecture~\cite{Nadin2010}, mathematical foundations of anticipatory behaviour~\cite{Dubois1998}, and real-world implementations of anticipatory computing in robotics~\cite{Stolzmann2000} have been achieved.} Yet the inability to seamlessly interact with the environment and sense feedback that will guide anticipatory learning is the major obstacle for further proliferation of anticipatory computing applications.


\subsection{Architecture of Anticipatory Mobile Systems}
\label{sec:architecture_anticipatory}

\begin{figure}[t]
    \centering
    \includegraphics[width=0.7\textwidth]{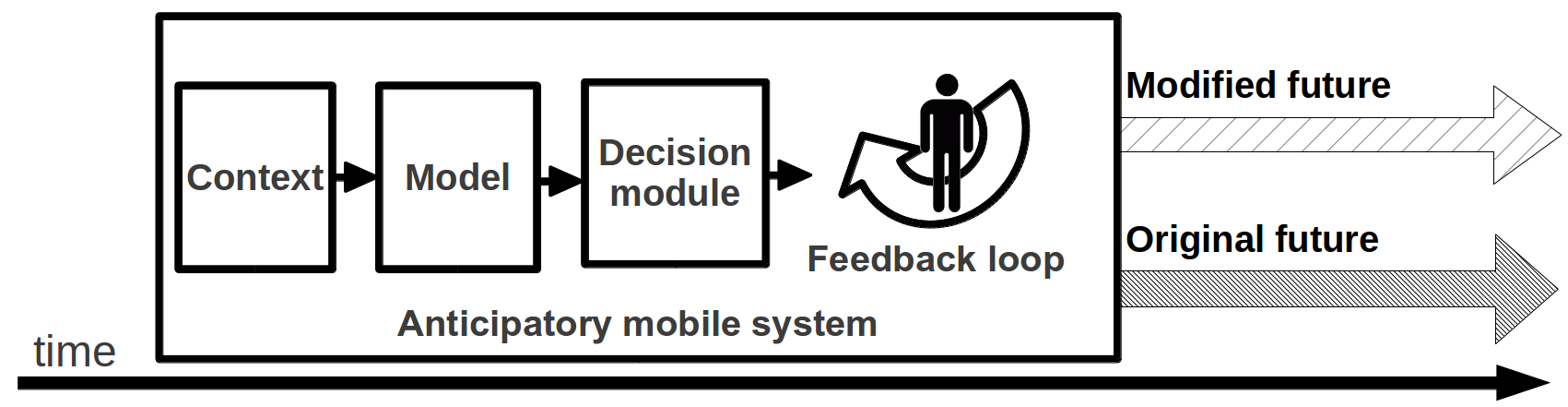}
    \caption{Anticipatory mobile systems predict context evolution and the impact their actions can have on the predicted context. The feedback loop consisting of a mobile and a human, enables the system to affect the future.}
 	\label{fig:feedback_loop}
\end{figure}

With smartphones the restriction on the interaction is lifted. Multimodal sensing and high processing capabilities of modern phones enable momentary awareness of the surrounding environment. At the same time, phones' anytime-anywhere use and a rich interface with the user enable a tight feedback loop ensuring that anticipatory decisions are realised. The symbiosis of the smartphone and the user allows for a new kind of a system -- \textit{anticipatory mobile computing system}. \red{In Figure 2 we adapt Nadin's conventional anticipatory computing architecture~\cite{Nadin2010} to mobile system design, sketching the anticipatory mobile system's core functional parts.} First, the surrounding context is sensed, then a predictive model of the context is built. At this point it is worthwhile to note the difference between a \textit{predictive} and an \textit{anticipatory} system. A predictive system has a model of what the future state of the context and/or the system itself will be. If the stress app presented in the previous section were merely predictive, it would predict the user's expected stress level and inform the user about it. An anticipatory system makes intelligent decisions in order to impact the future to the benefit of the user. Thus, a fully anticipatory version of our stress relief app would, after predicting dangerous stress levels, reschedule user's meetings according to the learnt model of stress evolution in order to improve user's well-being. In Figure~\ref{fig:feedback_loop} the decision module uses predicted future as a basis for deciding on system's actions. The action is selected so that it results in a favourable change in the future state of the system or the environment. The action is, in general, performed by the user who is influenced by the information provided by the smartphone. The phone remains in a feedback loop with the user: besides informing the user, the phone observes the outcome of its suggestions on the evolving model of the system. 

\begin{figure}[t]
    \centering
    \includegraphics[width=0.7\textwidth]{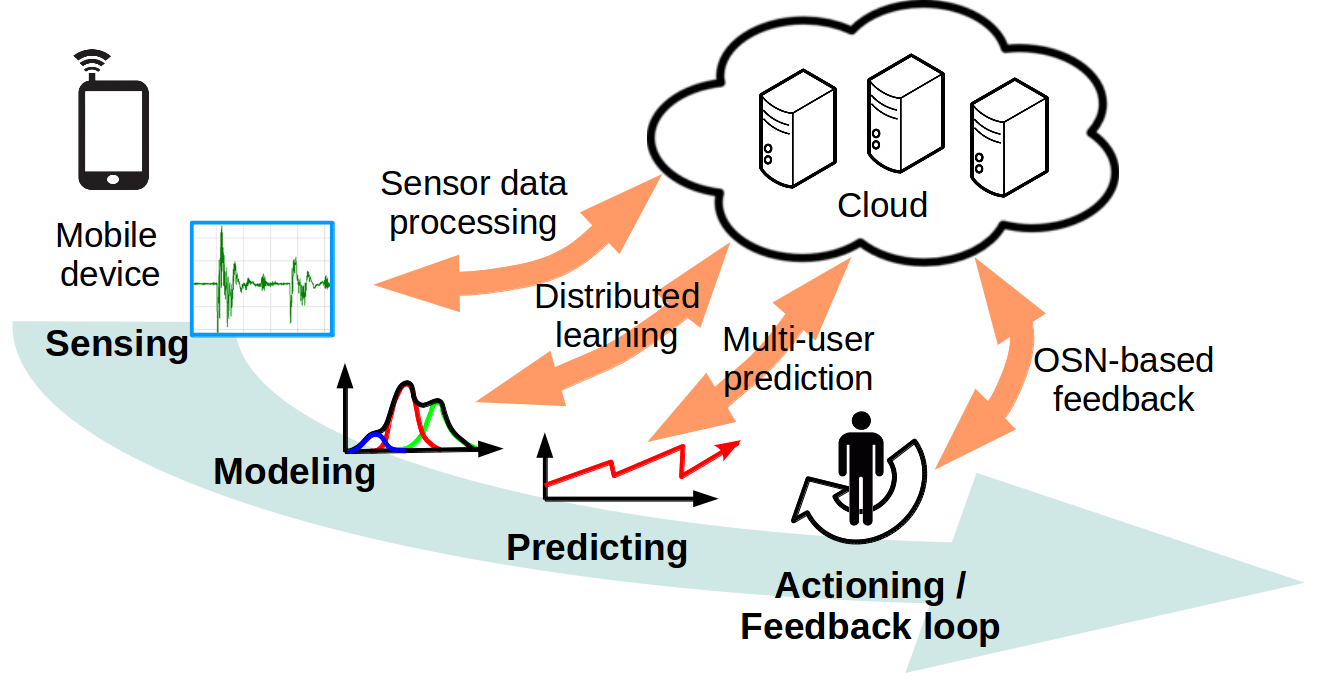}
    \caption{Anticipatory mobile computing architecture. The mobile senses, models and predicts the context, and through interaction with the user ensures that anticipatory decisions are implemented. At each step, the computation can be distributed between the mobile and the cloud.}
    \label{fig:anticipatory_architecture}
\end{figure}

Anticipatory mobile computing requires multiple processing stages, relationships among which are shown in Figure~\ref{fig:anticipatory_architecture}. The stages include context sensing and modelling, context prediction and impacting the future through interaction with the user. Unlike previously attempted anticipatory computing realisations, the proposed architecture can benefit from devices' always on connectivity. Thus, the phone can offload computation to the cloud, integrate predictions of multiple users in order to build more accurate models of context evolution, and can harness the power of online social networks for enhanced interaction with users.

\section{Context Sensing and Modelling for Anticipatory Computing}
\label{sec:sensing_modelling}

Mobile sensing has grown from the need for computing devices that are truly integrated with the everyday life of individuals. This can happen only if the devices are cognisant of their environment. Situations and entities that comprise the environment are collectively termed \textit{context}. Context may have numerous aspects: geographical, physical, social, temporal, or organisational, to name a few. Context sensing aims at bridging physical stimuli sensed by the device's sensors, also known as \textit{modalities}, and high level concepts that describe a context. Smartphones have evolved from communication devices to perceptive devices capable of inferring the surrounding context. 

Mobile phone's ability to infer that its user is jogging~\cite{Miluzzo2008}, commuting to work, sleeping \cite{Lane2011b} or even feeling angry \cite{Rachuri2010} is enabled by two factors. First, modern day smartphones are provisioned with sophisticated sensors, as well as with communication and computation hardware. A today's phone hosts a touch-screen, GPS, accelerometer, gyroscope, proximity and light sensors, a high quality microphone and cameras. Multi-core processors and gigabytes of memory allow smartphones to locally handle a large amount of data coming from these senses and extract meaningful situation descriptors, while a range of communication interfaces, such as WiFi, Bluetooth, 4G/LTE, and a near-field communication (NFC) interface, allow distributed computation and data storage. The second key factor that enables phones to make high level inferences is the increasingly ubiquitous and personal usage of mobile phones. Nowadays, the majority of the world's population owns a mobile phone, and these phones are closely integrated with people's lifestyle. These devices are not only physically present with their owners for most of the day, but are also used for highly personal purposes such as organising meetings, navigation, online social networking and e-commerce. 

\begin{figure}[t]
    \centering
    \includegraphics[width=\textwidth]{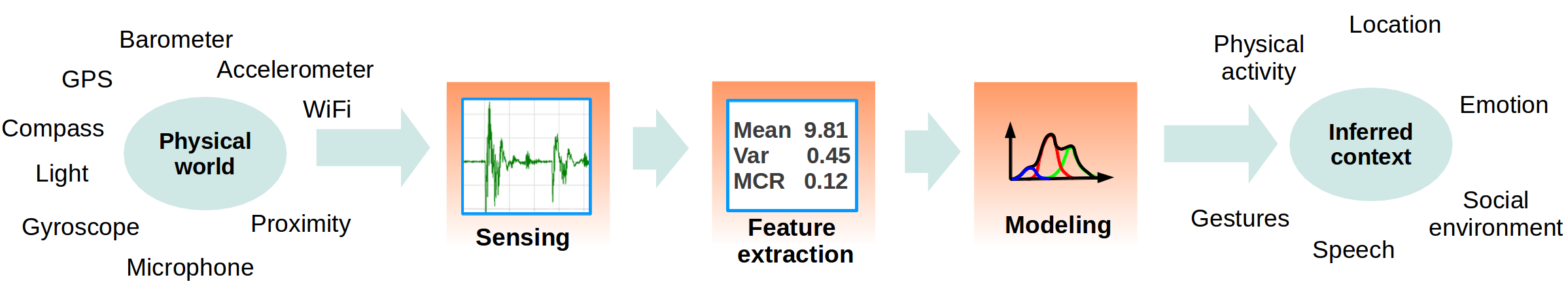}
    \caption{Mobile sensing: from real-world signals to high-level concepts.}
 	\label{fig:mobile_sensing_challenges}
\end{figure}

Context inference is a complex process that lies at the foundation of anticipatory mobile computing. Figure~\ref{fig:mobile_sensing_challenges} depicts the stages needed to get from environmental data to high-level inferences about the context. The first stage, \textit{sensing}, aims to provide an interface between the physical world and a mobile device. \textit{Feature extraction} is an intermediate step at which raw data are transformed to a form suitable for context inference. \textit{Modelling context} concentrates on the construction of models that connect interesting events or behaviours and extracted data features. 

\subsection{Multimodal Context Sensing for Anticipatory Mobile Computing}

Context sensing plays a major role in anticipatory mobile computing. First, sensed data serve as a basis for building predictive models of the phenomenon of interest. GPS tracking, for example, can be used to predict user whereabouts. \red{Second, mobile sensors can reveal high level information about users' internal state. In Section~\ref{sec:architecture_anticipatory} we note that future-changing actions in anticipatory mobile systems depend on the user to execute them. These actions can be communicated more efficiently, if the state of the user, such as user's mental load, attitudes and emotions, is known~\cite{Pejovic2014}.} 



A single sensor modality is seldom sufficient for inferring the context in which a device is. In addition, multimodal information can offset the ambiguities that arise when single sensor data are used for inference~\cite{Maurer2006}. Today's smartphones avail highly multimodal sensing, and are unobtrusively carried by their owners at all times. Beyond momentarily context inference, this also allows smartphones to sense multiple aspects of human behaviour, relate them, and uncover relationships previously unknown or difficult to confirm through conventional social science approaches. For example, mood can be correlated with user's location or activity~\cite{Puiatti2011},  socio-economic factors can be uncovered from calling and movement patterns\cite{Lathia2012,Martinez2012}, and mental and physical health can be assessed via mobile sensing \cite{Rabbi2011,MCMFP12:sensing}. In this section we pay particular attention to multimodal sensing for supporting anticipatory systems, and we put an accent on the affordances of, and challenges associated with smartphone based sensing.

\subsection{Implementation Issues}
\label{sec:sensing_challenges}


Applications that use smartphone sensing are subject to constraints coming from the devices' hardware restrictions. In anticipatory mobile computing frequent sensing of different modalities and collaboration of multiple agents are likely to be necessary for accurate anticipation, emphasising the need for resource-efficient mobile sensing solutions. Energy-efficient operation, processing, storage and communication constrains are the most common practical mobile sensing challenges. In Table~\ref{tab:challenges} we summarise the state-of-the-art solutions to address these issues.

\begin{table}
\caption{Context sensing challenges and possible solutions.}
\label{tab:challenges}
\begin{center}
\begin{scriptsize}
\begin{tabular}{|c|l|}
\hline
\textbf{Challenge} & \textbf{Solution} \\
\hline
\hline
\begin{tabular}[l]{@{}l@{}} Adaptation and \\context-driven operation\end{tabular} &
\begin{tabular}[l]{@{}l@{}} -- Adaptive sampling \cite{Kim2011,Rachuri2011}\\-- Hierarchical modality switching\\ \cite{Wang2009,Lu2010,Paek2010,Kang2008,Kim2010}\\-- Harnessing domain structure \cite{Foll2012,Nath2012,Paek2011}\\-- Cloud offloading \cite{Liu2012}
\end{tabular}\\
\hline
\begin{tabular}[l]{@{}l@{}} Computation, storage and \\ communication \end{tabular} &
\begin{tabular}[l]{@{}l@{}} -- Hierarchical processing \cite{Lu2010,Lee2013}\\
-- Cloud offloading\\ \cite{Miluzzo2008,Cuervo2010,Rachuri2011,Chun2011}\\
\red{-- Hardware co-processing} \cite{Priyantha2011,Lin2012}\end{tabular}\\
\hline
\end{tabular}
\end{scriptsize}

\end{center}
\end{table}

\subsubsection{Sensing Adaptation and Context-driven Operation}
Energy shortage issues are exacerbated by the design of smartphone sensors as occasionally used features, rather than constantly sampled sensors. Two popular means of reducing the energy consumption are adaptive sampling, i.e., sampling less often, and, in the case of a device with multiple sensors, powering them on hierarchically, i.e., preferring low-power sensors to more power hungry ones. 
In SociableSense, a mobile application that senses socialization among users \cite{Rachuri2011}, a linear reward-inaction function is associated with the sensing cycle, and the sampling rate is reduced during ``quiet" times, when no interesting events are observed. The approach is very efficient with human interaction inference, since the target events, such as conversations, are not sudden and short. On another side of the solution spectrum, the Energy Efficient Mobile Sensing System (EEMSS) proposed by Wang et al. hierarchically orders sensors with respect to their energy consumption, and activates high-resolution power-hungry sensors, only when low-consumption ones sense an interesting event~\cite{Wang2009}. Adaptive sampling and hierarchical sensing are not the only means of reducing energy usage. The inherent structure of the context inference problem can also be used to improve sensing efficiency. This is the main idea behind the Acquisitional Context Engine (ACE) proposed in \cite{Nath2012}. Here, Nath develops a speculation-based sensing engine that learns associative rules among contexts, an example of which would be ``when a user state is \textit{driving}, his location is not \textit{at home}". When a context-sensitive application needs to know if a user is is at home or not, it contacts ACE that acts as a middle layer between sensors and the application. ACE initially probes a less energy costly sensor -- accelerometer -- and only if the sensed data does not imply that a user is driving, it turns the GPS on and infers the actual user's location. While demonstrated on simple rules, Nath argues that ACE can be complemented with tools that examine the temporal continuity of context, such as SeeMon \cite{Kang2008} to extract sophisticated rules like ``if a user is \textit{at home} now, he cannot be \textit{in the office} in the next ten minutes". 

\subsubsection{Processing, Storage and Communication Efficiency} 

Despite ongoing technological advances mobile phones still have limited processing and data storage capabilities. Remote resources available via online cloud computing can be used to help with data processing. However, the transfer of the high-volume data produced by mobile sensors can be costly, especially if done via a cellular network. Balancing local and remote processing was tackled in one of the first smartphone sensing applications, CenceMe~\cite{Miluzzo2008}. This application performs audio and activity classification on the phone, while some other modalities, such as user's location, are classified on a remote server. The distribution of the computation is not performed solely because of the limited computation resources of a smartphone. Distributed computation also allows for aggregation of data from multiple phones, therefore a larger context can be inferred. \ignore{In CenceMe an ``Am I Hot" classifier of user's social behaviour is developed thanks to distributed, multi-user sensing. } In SociableSense~\cite{Rachuri2011}, the split between local and remote data processing is done on the basis of energy expenditure, data transmission cost, and the computation delay. 
Custom-made application execution partitioning, such as the one used in CenceMe and SociableSense, requires significant effort from the developer's side. More general solutions allow an application developer to delegate the partitioning task to a dedicated middleware. MAUI, for example, supports fine-grained code offloading to a cloud in order to maximise energy savings on a mobile device \cite{Cuervo2010}. 


\subsection{Context Modelling for Anticipatory Mobile Computing}
\label{sec:modeling_context}

Raw sensor data, such as those from phone's accelerometer, are seldom of direct interest and machine learning techniques are usually employed to infer higher level concepts, for example, a user's physical activity \cite{Tapia2007}. For the inference to be made, first we need to identify the most informative modalities and features of the raw sensor data, e.g. accelerometer data mean intensity and variance. Then, appropriate machine learning techniques are used to build a model of the phenomenon of interest, i.e. physical activity, and train the model with the data gathered so far.

\subsubsection{Selecting Useful Modalities and Features from Sensor Data}
\label{sec:select_features}
The first challenge in context modelling is the identification of those modalities of raw data that are the most descriptive of the context. Interdisciplinary efforts and domain knowledge are crucial in this step. For example, if we try to infer user's emotions, the existing work in psychology tells us that emotions are manifested in a person's speech \cite{Bezooijen1983}. Consequently, we can discard irrelevant modalities and concentrate our efforts on processing microphone data.

The next step includes the selection of the appropriate representation for the sensor data. Consider EmotionSense, an experimental psychology research application, which infers emotions from microphone data \cite{Rachuri2010}. Before the classification, however, raw microphone data has to be transformed into a suitable form. Distinctive properties extracted from the data are called \textit{features}. The EmotionSense authors built its speech recognition models on Perceptual Linear Predictive (PLP) coefficients, a well established approach to speech analysis \cite{Hermansky1990}. An alternative, but less sophisticated means of microphone data manipulation that has proven successful in mobile sensing is the Discrete Fourier Transform (DFT) \cite{Miluzzo2008}. The majority of speech energy is found in a relatively narrow band from 250 Hz to 600 Hz, thus the investigation of DFT coefficient means and variations can help identify speech in an audio trace. Finally, Mel-Frequency Cepstral Coefficients (MFCC) are another commonly used feature for speech recognition \cite{Lu2009,Miluzzo2010,Chon2012}.

\begin{table}[t]
\caption{Context sensing domains and characteristic features.}
\label{tab:features}
\begin{center}
\begin{scriptsize}
\begin{tabular}{|c|l|}
\hline
\textbf{Domain} & \textbf{Characteristic Features} \\
\hline
\hline
Speech recognition & 
\begin{tabular}[l]{@{}l@{}} -- Sound spectral entropy, RMS, zero crossing rate, low energy frame rate,\\
 spectral flux, spectral rolloff, bandwidth, phase deviation\\
\cite{Lu2009,Lane2011b}\\
-- Mel-Frequency Cepstral Coefficients (MFCCs)\\
\cite{Lu2009,Miluzzo2010,Chon2012,Lu2010,Lu2012}\\
-- Teager Energy Operator (TEO), pitch range, jitter and standard deviation,\\
spectral centroid, speaking rate, high frequency ratio\\
\cite{Lu2012,Lu2010}\\
-- Running average of amplitude, sum of absolute differences\\
\cite{Krause2006}\\
-- Perceptual linear predictive (PCP) coefficients\\
\cite{Rachuri2010}\\
-- Mean and standard deviation of DFT power\\
\cite{Miluzzo2008}


\end{tabular}\\
\hline
Activity classification & 
\begin{tabular}[l]{@{}l@{}}-- Accelerometer FFT principal component analysis (PCA)\\
\cite{Krause2006}\\
-- Accelerometer intensity/energy/mean\\
\cite{Eston1998,Lu2010,Rabbi2011,Abdullah2012}\\
-- Accelerometer variance\\
\cite{Miluzzo2008,Lu2010,Rabbi2011,Aharony2011}\\
-- Accelerometer peaks/mean crossing rate\\
\cite{Miluzzo2008,Lu2010}\\
-- Accelerometer spectral features\\
\cite{Lu2009,Lu2010,Lane2011b,Rabbi2011}\\
-- Accelerometer correlations\\
\cite{Abdullah2012}\\
-- Accelerometer frequency domain entropy\\
\cite{Abdullah2012}\\
-- Barometric pressure\\
\cite{Rabbi2011}

\end{tabular}\\
\hline

Location & 
\begin{tabular}[l]{@{}l@{}}
-- Days on which any cell tower was contacted,
days on which a specific tower is contacted, \\
contact duration, events during work/home hours\\
\cite{Isaacman2011}\\
-- Tanimoto Coefficient of WiFi fingerprints \\
\cite{Chon2012}\\
-- Eigenbehaviors - vectors of time-place pairs\\
\cite{Eagle2009b}\\
-- Hour of day, latitude, longitude, altitude, social ties\\
\cite{DeDomenico2012}\\
\end{tabular}\\
\hline
Object recognition & \begin{tabular}[l]{@{}l@{}}-- GIST features\\
\cite{Chon2012}
\end{tabular}\\
\hline
Gestures & \begin{tabular}[l]{@{}l@{}}-- Mean, max, min, median, amplitude, and
high pass filtered values of acc. intensity and jerk;\\
spectral features; screen touch location, slope, speed,
strokes number, length, slope and location\\
\cite{Coutrix2012}\\
\end{tabular}\\
\hline
Physiological state & \begin{tabular}[l]{@{}l@{}}-- Galvanic skin response, heat flux,\\
skin thermometer running average and sum of absolute differences\\
\cite{Krause2006}\\
\end{tabular}\\
\hline
Thoughts & \begin{tabular}[l]{@{}l@{}}-- Bandpass filtered neural signal from EEG\\
\cite{Campbell2010}\\
\end{tabular}\\
\hline
Call prediction & \begin{tabular}[l]{@{}l@{}}--  Call arrival and inter-departure time, calling reciprocity\\  \cite{Phithakkitnukoon2011}\\
\end{tabular}\\
\hline
Interruptibility & \begin{tabular}[l]{@{}l@{}}-- User typing, moving, clicking, application focus, app. activity, gaze, time of day,\\
day of week, calendar, acoustic energy, WiFi environment,\\
\cite{Horvitz2003}\\
-- Mean, energy, entropy and correlation of accelerometer data\\
\cite{Ho2005}
\end{tabular}\\
\hline
\end{tabular}
\end{scriptsize}

\end{center}
\end{table}

The above example shows that numerous features can be extracted from a single modality. In many domains, however, certain feature types have crystallised out as the most informative. Table~\ref{tab:features} lists the most commonly observed features and the domains in which they are used. The table is not meant to be a comprehensive survey of feature extraction, but should point out that even with a small number of sensors there can be hundreds of possible features all of which may or may not contribute to context inference~\cite{Choudhury2008}.  
Modality and feature selection impact the rest of the context inference; a careful consideration at this stage of the process can help improve classification accuracy or reduce the computational complexity of the learning process. As mobile sensing matures the variety of context types that we strive to infer broadens. In addition, the number of sensors available on the smartphone increases steadily. Therefore, identifying and quantifying the strength of a link between a domain and a modality (or a feature) emerges as an important research direction in mobile sensing.

\subsubsection{Classification Methods}
\label{sec:classification_methods}
A plethora of machine learning techniques can be used to transfer distilled sensor data into mathematical representations of a phone's environment or user's behaviour. In this survey we concentrate on a small subset of techniques that have been successfully applied in practice, and we refer an interested reader to machine learning texts such as~\cite{Bishop2006,elementsstatisticallearning2009,RG11:first,Bar12:Bayesian}.

We examine how context inference models are built in the case of StressSense, a mobile phone application that analyses speech data collected via a built-in microphone and identifies if a user is under stress \cite{Lu2012}. The first step in StressSense is sound and speech detection. The application assumes that sound is present if high audio level is detected in at least 50 out of 1000 samples taken within a half a second period. In such a case, StressSense divides the audio signal into frames and for each of the frames calculates its zero crossing rate (ZCR) and root mean square (RMS) of the sound. These features correspond to sound pitch and energy. A tree-based classifier that decides between speech and non-speech frames is built with ZCR and RMS as attributes. Further, thresholds on ZCR and spectral entropy are used to discern between voiced and unvoiced frames of human speech. Finally, Gaussian Mixture Models (GMMs) are built for the two target classes -- stressed speech and neutral speech. Pitch, Teager Energy Operator (TEO) and Mel-Frequency Cepstral Coefficient (MFCC) based features of each voiced frame are used for user stress inference.

The variety of classification methods and data features can be overwhelming for a mobile sensing application designer. To help with the selection of a context inference approach, in Table~\ref{tab:learners} we list mobile sensing challenges and the corresponding machine learning techniques that have been proved to work well in practice. The table is meant to be a starting point for mobile sensing practitioners, and does not imply that alternative techniques would not perform better. The structure of the problem at hand often hints towards an efficient classification approach. For example, Gaussian Mixture Models perform well when it comes to speaker identification, as it is possible to extract parameters for a set of Gaussian components
from the FFT of the speech signal and use them as a vectorial representation of human voice. This approach has been proved extremely effective for user identification~\cite{reynolds2000}. However, a deeper discussion about why certain approaches work in certain domains is outside of the scope of this survey.


\begin{table}
\caption{Context sensing domains and relevant machine learning techniques.}
\label{tab:learners}
\begin{center}
\begin{scriptsize}
\begin{tabular}{|c|l|}
\hline
\textbf{Domain} & \textbf{Machine learning technique} \\
\hline
\hline
Speech recognition & 
\begin{tabular}[l]{@{}l@{}} -- Hidden Markov Model (HMM)\\ \cite{Chon2012,Choudhury2003}\\-- Threshold based learning \cite{Wang2009}\\-- Gaussian Mixture Model (GMM) \cite{Rachuri2010,Lu2012}\end{tabular}\\
\hline
Activity classification & 
\begin{tabular}[l]{@{}l@{}}-- Boosted ensemble of weak learners\\ \cite{Consolvo2008,Abdullah2012}\\
-- Boosting \& HMM for smoothing \cite{Lester2005}\\-- Tree based learner \\  \cite{Tapia2007,Abdullah2012}\\-- Bayesian Networks~\cite{Krause2006}\end{tabular}\\
\hline
Location determination (with GPS) & 
\begin{tabular}[l]{@{}l@{}}-- Markov chain \cite{Ashbrook2003}\\
-- Non-linear time series \\ \cite{Scellato2011,DeDomenico2012}\end{tabular}\\
\hline
Location (with BT or WiFi) & \begin{tabular}[l]{@{}l@{}} -- Bayesian network\\
\cite{Eagle2006,Eagle2009a}\end{tabular}\\
\hline
Location (with ambient sensors) & -- Nearest neighbour \cite{Maurer2006}\\
\hline
Scene classification & -- K-means clustering \cite{Chon2012}\\
\hline
Object recognition & \begin{tabular}[l]{@{}l@{}}-- Support vector machine \cite{Chon2012}\\
-- Boosting \& tree-stump \cite{Wang2012}\end{tabular}\\
\hline
Place categorization & -- Labelled LDA~\cite{Chon2012}\\
\hline
Call prediction & -- Naive Bayesian \cite{Phithakkitnukoon2011}\\
\hline
Interruptibility & \begin{tabular}[l]{@{}l@{}}-- Bayesian Network \\ \cite{Horvitz2003,Fogarty2005}\\ -- Tree based learner \cite{Fogarty2005}\\
-- Naive Bayes~\cite{Hofte2007}\end{tabular}\\
\hline
\end{tabular}
\end{scriptsize}

\end{center}
\end{table}

\subsubsection{Handling Large-Scale Inference}
\label{sec:large_scale}


Anticipatory mobile computing applications for healthcare and personal assistance we sketched in Section~\ref{sec:anticipatory_glance} are of broad interest. We envision a multitude of such applications to be distributed through commercial app stores such as Apple App Store and Google Play. Scaling up the number of users imposes novel challenges with respect to sensing application distribution, data processing and scalable machine learning. Data diversity calls for more complex classification: walking performed by an eighty year old person will yield significantly different accelerometer readings than when the same activity is performed by a twenty year old. Clearly, classification needs to be less general, but does that imply a personal classifier for each user? 

In~\cite{Lane2011a} the authors propose community similarity networks (CSNs) that connect users who exhibit similar behaviour. User alikeness is calculated on the basis of their physical characteristics, their lifestyle and from the similarity between their smartphone-sensed data. For each of these three layers in the CSN, a separate boosting-based classifier is trained for any individual user. However, a single-layer classifier is trained on the data coming from not only the host user, but also from all the other users who show strong similarity on that CSN layer. In this way the CSN approach tackles the shortage of labelled data for the construction of personalised models, a common issue in large-scale mobile sensing.

Besides increased user diversity, mobile sensing applications interested in monitoring user behaviour often have to cope with long-term observations. In their ``social fMRI'' study Aharony et al. continuously gather over 25 sensing modalities for more than a year from about 130 participants \cite{Aharony2011}. Machine learning algorithms need explicit labelling of the high level concepts that are extracted from sensor readings. However, with highly multimodal sensing integrated with everyday life, querying users to provide descriptions of their activities becomes an intrusive procedure that may annoy them. Instead, a semi-supervised learning technique called \textit{co-training} is used to establish a bond between those sensor readings for which labels exist, and those for which only sensor data are present~\cite{Zhu2009}. Co-training develops two classifiers that provide complementary information about the training set. After the training on the labelled data, the classifiers are iteratively run to assign labels to the unlabelled portion of the data. In the mobile realm, unlabelled data representing an activity of one user could be similar to labelled data of the same activity performed by another user. In this case, labels can propagate through the similarity network of users~\cite{Abdullah2012}. 

In addition to a larger user base and an increased amount of gathered data, mobile sensing is further challenged by a growing number of devices used for context-aware applications. We increasingly observe \textit{ecosystems of devices}, where multiple devices work together towards improved context sensing. Fitbit, for example, markets a range of wearable devices that track user metrics such as activity, sleep patterns, and weight \cite{Fitbit}. As the popularity of these devices grows we can expect that a single user will carry a number of context sensing devices. Darwin Phones project tackles distributed context inference where multiple phones collaborate on sensing the same event \cite{Miluzzo2010}. First, via a cloud infrastructure, phones exchange locally developed models of the target phenomenon. Later, when the same event is sensed by different phones, inference information from each of the phones is pulled together so that the most confident description of the event is selected.

In this section we summarised how machine learning can be used for context inference. Machine learning techniques are crucial for context prediction and anticipatory decision making, two other steps of anticipatory computing. Unlike context inference, these two areas are less explored. Their real-world implementations are scarce, and in the following sections we present recent advances in mobile prediction and anticipatory decision-making. Integration of machine learning approaches in context inference, prediction and anticipation, however, remains an interesting research challenge.




\section{Context Prediction}
\label{sec:context_prediction}


Predictions of human behaviour, crucial for many anticipatory computing applications, are for the first time available to application developers. These predictions are enabled by the close integration of the phone and the user, which allows the phone to record user's context at all time, and the fact that humans remain creatures of habit and patterns of behaviour can be identified in the sensed data.

\subsection{Mobility Prediction}
\label{sec:mobility_prediction}
Historically, the prediction of mobile phone users' movement patterns was tied with system optimisations. Anticipation of surges in the density of subscribers in a cellular network was proposed for dynamic resource reservation and prioritised call handoff~\cite{Soh2003}. Yet, as data collected by a phone gets more personal, the opportunity for novel user-centric applications increases. User movement can be examined on different scales. For small-scale indoor movement predictions, systems can rely on sensors embedded in the buildings. An example of such systems is MavHome: the authors proposes a smart home which adjusts indoor light and heating according to predicted movement of house inhabitants~\cite{Cook2003}. A large part of the current research, however, concentrates on the city-scale prediction of users' movement. In addition, predicted location can be considered on a level higher than geographical coordinates. Work of Ashbrook and Starner, as well as of Hightower et al. aims to recognise and predict places that are of special significance to the user~\cite{Ashbrook2003,hightower2005}. The interest in such prediction was further raised with proliferation of smartphones and commercial location-based services such as Foursquare \cite{Foursquare}. In such as setting, targeted ads can be disseminated to phones of users who are expected to devote a certain amount of their time to eating out or entertainment. The NextPlace project aims to predict not only user's future location, but also the time of arrival and the interval of time spent at that location \cite{Scellato2011}. The authors base the prediction on a non-linear time series analysis. More recently, Horvitz and Krumm devised a method for predicting a user's destination and suggesting the optimal diversion should the user want to interrupt his/her current trip in order to, for example, take a coffee break \cite{Horvitz2012}. Noulas et al. investigate next check-in prediction in the Foursquare network \cite{Noulas2012}. They show that a supervised learning approach that takes into account multiple features, such as the history of visited venues, their overall popularity, observed transitions between place categories, and other features, is needed for successful prediction. SmartDC merges significant location prediction with energy-efficient sensing, and proposes an adaptive duty cycling scheme to provide contextual information about mobility of users~\cite{Chon2013}.  





Research on mobility prediction was additionally boosted by large sets of multimodal data that has been collected and made publicly available by companies and academic institutions. For example, the MIT Reality Mining~\cite{Eagle2006} project accumulated a collection of traces from one hundred subjects monitored over a period of nine months. Each phone was preloaded with an application that logged incoming and outgoing calls, Bluetooth devices in proximity, cell tower IDs, application usage, and phone charging status. Similarly, the Nokia Mobile Data Challenge (MDC) data set was collected from around 200 individuals over more than a year~\cite{Laurila2012}. The logs contain information related to GPS, WiFi, Bluetooth and accelerometer traces, but also call and SMS logs, multimedia and application usage. The above data sets served as a proving ground for a number of approaches towards mobility prediction. In~\cite{Eagle2009a} Eagle et al. demonstrate the potential of existing community detection methodologies to identify significant locations based on the network generated by cell tower transitions. The authors use a dynamic Bayesian network of places, conditioned on towers, and evaluate the prediction on the Reality Mining data set. De Domenico et al. exploit movement correlation and social ties for location prediction~\cite{DeDomenico2012}.  Relying on nonlinear time series analysis of movement traces that do not originate from the user, but from user's friends or people with correlated mobility patterns, the authors demonstrate improved accuracy of prediction on the MDC data set. Interdependence of friendships and mobility in a location-based social network was also analysed in~\cite{Cho2011}. \red{McInerney et al. propose a method, based on a novel information-theoretic metric called instantaneous entropy, for predicting departures from routine in individual's mobility~\cite{McInerney2013}. Such predictions are of extreme importance for personalised anticipatory mobile computing applications, for example the ones that aim to elicit a positive behaviour change in a human subject~\cite{Pejovic14:anticipatory}.}

\red{Different approaches to mobility prediction make different assumptions about human mobility. Markov predictors often assume that people spend similar residence time at the same places, while non-linear time series approaches assume that people spend similar staying time at similar times of a day \cite{Chon2013}. Additionally, certain real-world restrictions, such as the fact that ground movement has to follow the road network, can figure in prediction methods. In Table~\ref{tab:mobility_prediction} we list, and provide examples of, commonly used mobility prediction methods.}



\begin{table}
\caption{\red{Modelling Methods for Mobility Prediction.}}
\label{tab:mobility_prediction}
\begin{center}
\begin{scriptsize}
\begin{tabular}{|c|l|}
\hline
\textbf{Method} & \textbf{Example} \\
\hline
\hline
\begin{tabular}[l]{@{}l@{}} Markovian \end{tabular} &
\begin{tabular}[l]{@{}l@{}} -- Markov process (MP) \cite{Ashbrook2003,Song2004}
\end{tabular}\\
\hline
\begin{tabular}[c]{@{}c@{}} Nonlinear time \\series analysis (NTSA) \end{tabular} &
\begin{tabular}[l]{@{}l@{}} -- NTSA \cite{Scellato2011}\\
-- NTSA with social information \cite{DeDomenico2012}\end{tabular}\\
\hline
\begin{tabular}[l]{@{}l@{}} Bayesian \end{tabular} &
\begin{tabular}[l]{@{}l@{}} -- Dynamic Bayesian Network \cite{Eagle2006,Eagle2009a}\\\cite{McInerney2013,Etter2013}\\
-- Road-topology-aware with Bayes rule \cite{Ziebart08}\\
\end{tabular}\\
\hline
\begin{tabular}[l]{@{}l@{}} Other/Hybrid \end{tabular} &
\begin{tabular}[l]{@{}l@{}} -- MP with NTSA \cite{Chon2013}\\
-- Road-topology-aware MP \cite{Soh2003}\\
-- Information-theoretic uncertainty minimisation \\\cite{Bhattacharya2001,Cook2003} \\
-- Probabilistic road-topology aware \cite{Horvitz2012}\\
-- Statistical regularity-based model \cite{McNamara2008}, \\
-- Temporal, spatial and social probabilistic model  \cite{Cho2011},\\
-- Frequent meaningful pattern extraction \cite{Sadilek2012}\\
-- M5 trees and linear regression \cite{Noulas2012}
 \end{tabular}\\
\hline
\end{tabular}
\end{scriptsize}
\end{center}
\end{table}

\subsection{Lifestyle, Health and Opinion Prediction}

Multimodal traces also enable prediction of behavioural aspects beyond mobility. Human activity prediction, for example, has been an active subject of research in the past years: various approaches have been presented in the literature, based for example on accelerometers~\cite{Choudhury2008,Tapia2007}, state-change sensors \cite{Tapia2004} or a system of RFIDs \cite{Wyatt2005}. In a series of seminal works such as~\cite{LFK05:Liao,liao2007hcr}, Liao et al. demonstrate the prediction and correlation of activities using location information. Eagle and Pentland propose the use of multimodal \textit{eigenbehaviours}~\cite{Eagle2009b} for behaviour prediction. Eigenbehaviours are vectors that describe key characteristics of observed human behaviour over a specified time interval, essentially lifestyle. The vectors are obtained through the principal component analysis (PCA) of a matrix that describes a deviation in sensed features. Besides being a convenient notation for time-variant behaviour, by means of simple Euclidean distance calculation, eigenbehaviours enable direct comparison of behaviour patterns of different individuals. Eagle and Pentland demonstrate the ability of eigenbehaviours to recognise  structures in behaviours by identifying different groups of students at MIT.

Certain aspects of the context that are internal to the user can also be predicted. In their work on health status prediction, Madan et al. use mobile phone based co-location and communication sensing to measure characteristic behaviour changes in symptomatic individuals~\cite{Madan2010}. The authors find that health status can be predicted with such modalities as calling behaviour, diversity and entropy of face-to-face interactions and user movement patterns. Interestingly, they demonstrate that both physiological as well as mental states can be predicted by the proposed framework. Our running example of an anticipatory stress relief app could rely on such internal well-being state predictions. Finally, political opinion fluctuation is a topic of another work by Madan et al.~\cite{Madan2011a}, which shows the potential use of the information collected via mobile sensing for understanding and predicting human behaviour at scale. In this work call and SMS records, Bluetooth and WiFi environment are used to model opinion change during the 2008 presidential elections in the United States. Face-to-face personal encounters, measured through Bluetooth and WiFi collocation, are the key factor in opinion dissemination. The authors also discover patterns of dynamic homophily related to external political events, such as election debates. Automatically estimated exposure to a political faction can predict individual's opinion on the election day.


\section{Closing the Loop: Shaping the Future with Anticipatory Computing}
\label{sec:closing_loop}

Theoretical underpinnings of anticipatory computing have been laid down in the last few decades. Practical applications are lacking due to inability to maintain tight interaction of a computing system, its environment and a user. Smartphones for the first time enable a quick \textit{model -- action -- effect} feedback loop for anticipatory computing.
\subsection{Persuasive Mobile Computing}

The existence of the feedback loop can be observed on the example of digital behaviour change intervention (dBCI) applications. These applications harness a unique perspective that a personal device has about its user to catalyse positive behavioural change. Behaviour change can address some of the most prevalent health and well-being problems, including obesity, depression, alcohol and tobacco abuse. Delivered via smartphones dBCIs support those who seek the change with timely and relevant information about the actions that should be taken. With smartphones, interventions scale to a potentially very large number of users, and can be delivered in accordance to user's momentarily behaviour and state. 

UbiFit~\cite{Consolvo2008} and BeWell~\cite{Lane2011b}, although not behavioural interventions in the strict therapeutic sense, represent the first step towards mobile dBCIs. In the former a phone's ambient background displays a garden that grows as user's behaviour gets in accordance with predefined physical activity goals. In BeWell, core aspects of physical, social, and mental well-being -- sleep, physical activity, and social interactions -- are monitored via phone's built-in sensors. For example, sleep patterns are inferred from phone recharging events and periods when a phone's microphone indicates near-silent environment. The feedback is provided via a mobile phone ambient display which shows an aquatic ecosystem where the number and the activity of animals depend on user's well-being.  Among the early dBCI  applications we find SociableSense, an app that examines the socialisation network within an enterprise and provides feedback about individual sociability~\cite{Rachuri2011}. Similarly, SocioPhone monitors turn taking in face-to-face interactions and enables dBCI applications to be designed on top of it~\cite{Lee2013}. One of the applications proposed by the authors is SocioTherapist. Designed for autistic children, SocioTherapist presents a game in which a child is rewarded each time it performs a successful turn taking. Social environment is also used as a motivator in the Social fMRI, an application that aims to increase physical activity of its users~\cite{Aharony2011}. In Social fMRI a close circle of friends get automatic updates whenever an individual phone registers that its user is exercising, promoting a competitive and stimulating environment. 

It is interesting to note that the mobile phone is the most personal computing device people have. Feelings that the users have towards their phones parallel those that they have towards their fellow humans~\cite{Lindstrom2011}. The above examples show that this relationship can be harnessed for influencing users' behaviour, bringing us to the concept of~\textit{persuasive mobile sensing}~\cite{Lane2010}. What remains unclear is the most appropriate modality of mobile-human interaction. Indeed, UbiFit and BeWell exploit innovative user interface techniques to close the loop between mobile sensing and actionable feedback. The ambient display is always present, and each time a phone is used, its owner gets a picture of his or her physical activity and level of sociability. For many other applications that need to deliver an explicit timely advice, interaction with the user is an open problem: if the user has to be notified via SMS, for example, how often should a message be sent, at what time, in which context? These are typical human-computer interaction questions related to interruptibility. 

\subsection{Personalised Interaction}


Smartphone's ability to sense and predict the user's context can serve as a basis for interaction adaptation and seamless integration with the user's daily routine. In his 1991 manifesto of ubiquitous computing Weiser advocates pervasive technology that coexists unobtrusively with its users \cite{Weiser1991}. This ``calm technology" is not our current reality, and indeed we get an abundance of notifications from an increasing number of devices we own. Thus, we receive irrelevant instant messages while working on an important project, a phone may sound an embarrassing ``out of battery" tone in the middle of a meeting, and a software update pop-up may show up while we are just temporarily connected to a hot spot in a coffee shop. From the anticipatory mobile computing point of view, inappropriate interaction moments potentially reduce the ability to impact the future with current actions, as the user, annoyed by the poorly communicated information, may decide to ignore it.


Attentive user interfaces manage user attention so that the technology works in symbiosis with, rather than against user's interruptibility. Context sensors proved to be instrumental in identifying opportune moments to interrupt a user. Performed before the smartphone era, early experiments relied on external sensors, such as a camera and a microphone, along with the information about user's desktop computer usage~\cite{Horvitz2003}. Horvitz et al. developed a framework for inferring user's workload in an office setting via a Bayesian network in which variables such as the presence of voice, user's head position and gaze, and currently opened applications on a user's PC are connected with the probability distribution of interruptibility. The idea of connecting sensed data with user interruptibility was reconsidered with early mobile computing devices. Ho and Intille investigate the interruption burden in case of mobile notifications~\cite{Ho2005}. Their study uses on-body accelerometers, and triggers interruptions only when a user switches her activity. The authors find that moments of changing activity, as inferred by the accelerometers, represent times at which an interruption results in minimal annoyance to the recipient. Fischer et al. demonstrate that interruptions coming immediately after the episodes of mobile phone activity, such as a phone call completion or a text message sending event, result in a more responsive user behaviour~\cite{Fischer2011}. Pielot et al. collected a data set of text messages exchanged via smartphones together with the associated phone usage context~\cite{Pielot2014}. Time since the screen was on, time since the last notification, and similar features were used in a classifier that infers if the users is going to attend the message within a short time frame. In~\cite{Pejovic2014} the authors discuss the design and implementation of InterruptMe, a real-time interruptibility inference framework that maintains a sensor data-based classifier of user interruptibility. The authors show that context, as sensed by a smartphone, can be used to identify moments when a user is likely to react to the delivered piece of information. 

\subsection{Online Social Networks for Anticipatory Actioning}

Although conceived as platforms for fun, leisurely interaction and information dissemination, Online Social Networks (OSNs) are increasingly being recognised for their persuasive power. Given their popularity, they might be used to influence the future behaviour of users. For example, through social reinforcement an individual's health-related behaviour is influenced by the behaviour of her OSN neighbours~\cite{Centola2010}. 
Moreover, in a controversial study on emotional expression on Facebook, Kramer et al. showed that emotions expressed by others in our OSN vicinity impact our own emotional expression~\cite{Kramer2014}. 

Anticipatory computing applications can use OSNs for both information dissemination tool, as well as for indirect persuasion. For example, an anticipatory traffic management application can send proactive driving directions via Twitter to a large number of users. Highly personalised mobile applications that aim to improve users' well-being, on the other hand, can harness social contagion to improve users' state. For example, obese people tend to have obese friends~\cite{Christakis2007} and a well-being application could prevent a user from becoming obese by proactively tackling obesity in the user's social circle. Although with a potential for high impact, OSN-based anticipatory behaviour intervention applications pose serious ethical challenges~\cite{Pejovic14:anticipatory}. The issues are exacerbated by the latent effect of OSN actions on users who are not even taking a part in the application.

\subsection{Anticipatory Decisions}

The timing and the means of information delivery are important for anticipatory actions to be picked up and performed by the user. Yet, the delivery becomes irrelevant if the action does not induce the preferred change in the future state. Deciding on the action is the core problem of anticipatory computing and a significant body of research deals with artificial implementations of anticipatory decision logic~\cite{Rosen1985,Butz2003}. In addition, two types of anticipatory behaviour are examined in the literature: implicit and explicit. Implicit anticipation refers to the case where decisions are embedded in the program of the system beforehand. Instead, explicitly anticipatory systems maintain a model of the environment and learn how to interact with the environment during their lifetime. We are particularly interested in explicit anticipation as we see it suitable for mobile sensing devices. 
A thorough discussion of anticipatory behaviour in adaptive learning systems, however, is beyond the scope of this survey, and for more details we refer an interested reader to~\cite{Butz2003b}. Instead, in the following we discuss some key implementation issues for a practical smartphone-based anticipatory mobile computing system.

\subsubsection{Reinforcement Learning} Mobile phones, carried by their owners at all times, are subject to frequent context changes that depend on the individual behaviour of the user. Therefore, pre-programmed implicit anticipation is unlikely to be feasible; for this reason, we concentrate on the explicit modelling of the context evolution. Such a model can be based on the types of predictions discussed in Section~\ref{sec:context_prediction}. The anticipatory decision module has to make a decision based on the predicted future. In case the problem space can be cast to the Markovian framework, i.e., if the current state depends only on the previous state, we can represent the state of the model and the rewards associated with each of the actions as a Markov Decision Process (MDP). Through reinforcement learning the system evaluates the reward it gets for an action performed in a certain state, with the goal of maximising the payoff~\cite{Sutton1998}. Just as biological systems learn from mistakes, so an artificial system reinforces actions that lead to favourable outcomes, and suppresses the others. To give a practical example, consider the stress relief smartphone application in Figure~\ref{fig:overview}. The application will occasionally reschedule user's meetings. Just like in the biological learning, the consequence will be evaluated, and if the application made a mistake, i.e., if the changes to the schedule turn out to be counterproductive, interfere with user's lifestyle or actually cause more stress for the user, a lesson will be learnt. However, two major issues arise with reinforcement learning in this situation. First, how does the application obtain signals that guide the learning? This can be done by an explicit query to the user, essentially asking the person if he is happy with decisions made by the application. Yet, the user may consider frequent querying to be irritating. Another option is to look for implicit signals. For example, monitoring if the user makes changes to the schedule immediately after the application interventions. The second issue comes from the intrinsic need of the reinforcement learning to make mistakes in order to learn from them. If the mistakes are costly (and increasing someone's stress level can surely be costly) the application should be careful about experimenting with decisions made with low confidence.  Thus, there is a trade-off, commonly known as \textit{the exploitation versus exploration trade-off}, between improving reinforcement learning models and minimising negative impact on the user \cite{Sutton1998}. 

\subsubsection{Learning without Interfering} To solve the problem of intrusive probing, we can employ \textit{latent learning}, a form of learning that takes place when a subject is immersed into an unknown environment without any rewards or punishments associated to the environment~\cite{Tolman1932}. Despite the lack of obvious incentives for learning, experiments with both humans and animals show that subjects form a cognitive map of the environment solely because they experience the world around them. Later, that cognitive map figures in decision making, essentially behaving as a learnt concept. The artificial implementation of latent learning has been demonstrated for example in~\cite{Stolzmann2000}. In this work the authors designed a robot that, just like in experiments with living rats, relies on latent learning to finds its way around a maze. Latent learning relies on the subject's ability to sense the environment. Immersed in the constant sensing of a large number of modalities, mobile phones can bring artificial latent learning to the next level. In the stress relief application example, instead of learning how to reschedule meetings only when a user gives feedback on the proposed schedule, a phone could passively monitor a user's meeting pattern, construct a latent model of events and take it into account even during the first rescheduling instance when the feedback is not yet available. Other approaches that are relevant for learning anticipatory systems are those belonging to the area of so-called probabilistic robotics~\cite{Thrun2005}.

Finally, so far we have proposed implementations of learning context evolution and action-reward modelling. The definition of anticipatory systems additionally calls for modelling the effect that actions will have on the context. While predicting the reward from the environment (i.e., its \textit{action value}) represents the key feature of reinforcement learning, a true anticipatory system should be able to predict the consequence of its interaction on the environment (i.e., the \textit{action effect})~\cite{Lanzi2008}. Recent progress in this area has been made through the development of \textit{anticipatory classifier systems} \cite{Stolzmann1998}.  In these systems the expectation of a future environment state is embedded within classifiers that model the problem. These classifiers are organised in a population that evolves over the course of its interaction with the environment. The evolving collection of classifiers itself is known as the \textit{learning classifier system}, and has been a subject of extensive research, an overview of which can be found in the survey article by Lanzi~\cite{Lanzi2008}.

\section{Towards Large-Scale Anticipatory Mobile Computing}
\label{sec:large_scale}

\begin{figure}[t]
    \centering
    \includegraphics[width=0.7\textwidth]{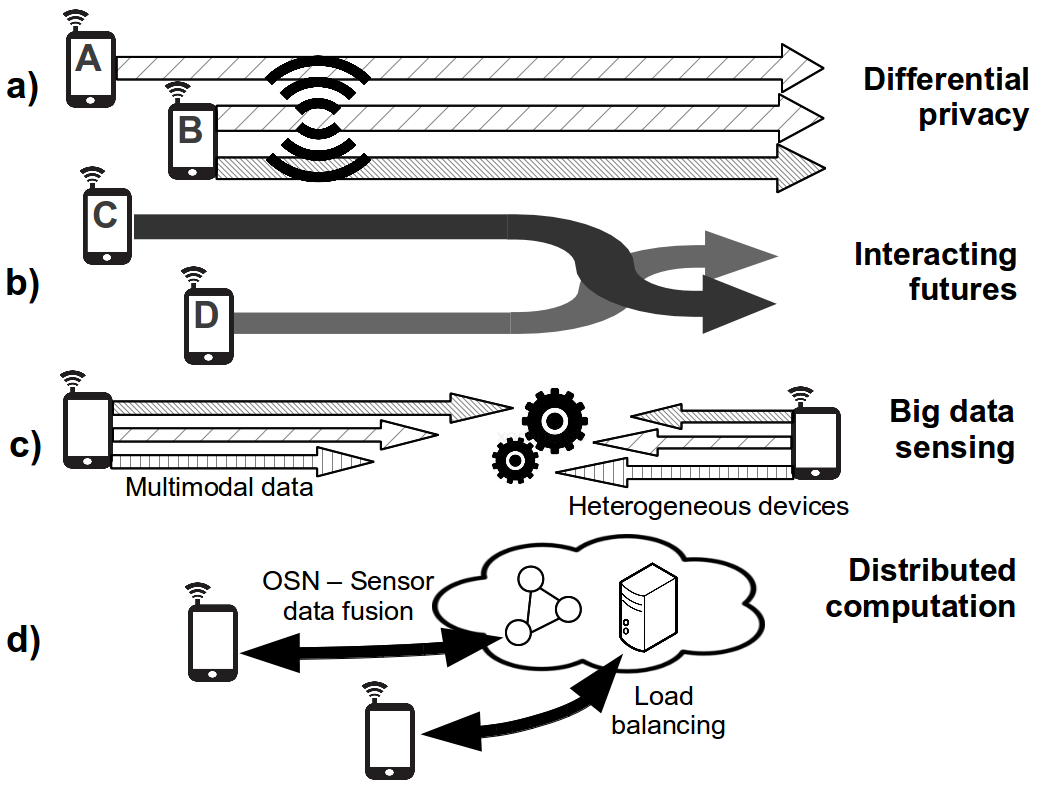}
    \caption{Large scale anticipatory computing challenges.}
    \label{fig:challenges_scale}
\end{figure}

Applications that rely on anticipatory computing are posed to be more human-like in their behaviour than legacy ones without predictive reasoning. Whether as personal assistants, doctors or even parents, anticipatory applications can provide domain-expert knowledge and personalised advising. Yet, these applications can go even further since, unlike humans, they are not constrained to a single subjective view of a situation. Rather, multiple phones can collaborate  towards common predictions and interactions. Such a large-scale anticipatory system introduces novel challenges (Figure~\ref{fig:challenges_scale}), which we summarise in the following subsections.

\subsection{Privacy and Anonymity} Smartphones can provide information about their owner's location, activities, and emotions. In addition, phones are, for most of the time, physically close to their users and connected to the global network. Thus, privacy and anonymity breaches in mobile sensing can result in unprecedented leaks of personal data. 
This information might be related not only to the current state of an individual, but also about the predictions made about her. For example, sensitive information might include future locations of an individual. 
Another privacy problem indigenous for mobile sensing stems from mismatched privacy policies over multiple users. We sketch the problem in Figure~\ref{fig:challenges_scale}.a). User $A$ volunteers his Bluetooth data only. User $B$ on the other hand, volunteers her GPS and her Bluetooth data. When users $A$ and $B$ are collocated, the information about user $A$'s location is inadvertently exposed. Although observable in general mobile sensing area, this privacy problem is exacerbated in anticipatory mobile computing. In fact, latent learning in an anticipatory computing application can include modelling the behaviour of a person who does not use the application himself. 
This is a classic problem of differential privacy~\cite{Dwork2006}: the new challenge here is to consider not only the current data about a certain individual, but also the information predicted about her. In anticipatory mobile computing, differential privacy is unlikely to be sufficient, since, as shown in the previous example, privacy of non-participating individual can be breached through latent sensing.


\subsection{Interacting Futures} High density of sensing devices, including smartphones, wearables and environment sensing networks, calls for joint consideration of all the sensing sources when it comes to anticipatory mobile computing. Frameworks for collaborative sensing include AnonySense~\cite{Cornelius2008}, Pogo~\cite{Brouwers2012} and PRISM~\cite{Das2010}. These projects allow application developers to distribute sensing over a large number of mobile devices and process individual devices' sensor data in a centralised way. In Section \ref{sec:large_scale} we already noted that context sensing can harness multiple views of the same event for improved inference. When it comes to context prediction and intelligent decision making, however, collaboration raises new research questions. Here multiple users do not only observe, but participate in context shaping. As sketched in Figure~\ref{fig:challenges_scale}.b), one person's future often depends on anticipated behaviour of another person. This interaction of subjective futures is evident on a simple example of traffic regulation. An intelligent phone might suggest an alternative route when a traffic congestion is anticipated on the usual route. Yet, if everyone involved in the congestion is advised to take the same alternative route, the conditions on which the route calculation is based will change, and the routing suggestions lose their value. 

\subsection{Big Data Sensing} We are living in a world characterised by enormous amounts of data. We post photos on Facebook, we tweet about our thoughts, and we carry smartphones that sense every second of our lives. Traditional methods of understanding human behaviour are insufficient to cope with the growing influx of personalised digital data \cite{Lazer2009}. Chon et al. mined opportunistically sensed multimodal data coming from multiple users in order to characterise places that those users visit~\cite{Chon2012}. While such a fusion of sensed data improves inference, a large amount of streaming data represent a challenge when it comes to storage and processing. Stream processing solutions have been proposed for centrally-managed static sensor networks~\cite{Madden2002}, but processing heterogeneous, opportunistically collected data remains an open problem. Overload with multiple sensing applications on a single phone, heterogeneity of smartphones, and high bandwidth demand resulting from data aggregation are identified as main challenges of large-scale crowdsensing~\cite{Xiao2013}. In addition, a large number of sources makes it hard to monitor the quality of a single source. Sensor malfunctioning, malicious participants or the fact that human behaviour might change when the subjects are aware of the sensors~\cite{Davis2012}, can lead to erroneous context inference and prediction.

\subsection{Distributed Computation} Distributed processing can be used at any stage of the anticipatory mobile computing pipeline, for improving resource utilisation~\cite{Rachuri2011} or context inference accuracy~\cite{Miluzzo2010}. The cloud also enables online social network and sensor data fusion~\cite{Yerva2012}, which can lead to richer context inference and novel context-aware OSN applications. Distributed computation does not come for free: communication cost, unreliable nodes and delay in gathering results are some of the main obstacles. In addition, OSN-data fusion is challenging due to distributed coordination of OSN querying and mobile sensing. Interestingly, anticipatory computing can both benefit and improve distributed processing. The prediction itself of user connectivity, presented in~\cite{Nicholson2008}, can guide decisions on whether to process sensed data locally or in the cloud. Utilisation of cloud resources or other devices in the future vicinity of the users (see \textit{cyber foraging}~\cite{Satyanarayanan2001}) can also be anticipated and exploited in the distribution of the computational load. 


\section{Challenges and Opportunities}
\label{sec:challenges}

It is possible to identify several research challenges that need to be addressed in order to realise the vision of anticipatory mobile computing. We now discuss some key topics, which in our opinion represent both challenges and opportunities for the research and industrial communities in the years to come.

\subsection{\red{Implementing Anticipatory Mobile Computing Systems}}

\subsubsection{\red{Non-deterministic behaviour in computing systems}}
Anticipation is unlikely to be deterministic. Predictions are done with a certain level of confidence, and biological systems often consider multiple possible futures in parallel. Nadin augments Rosen's definition of anticipatory system with the idea of unpredictability of the future, i.e., many possible futures might be possible given a current situation, and argues that a non-deterministic computer has to be at the core of an anticipatory system~\cite{Nadin2010}. Implementation of such a system with a smartphone, a deterministic machine with a limited ability to fully mimic biological systems, seems challenging. We argue, however, that its tight bond with the user makes the smartphone an ideal platform for anticipatory computing. While a device has a capability of bringing autonomous decisions based on its internal models of context evolution, actions are for most part taken in accordance with the user, possibly step by step. The user is guided by the phone, yet she considers the whole spectrum of possibilities before performing the action. Anticipatory applications might need a varying level of autonomy, and not all of them benefit from user's direct involvement in actioning. How, and whether at all, can we implement non-deterministic behaviour in a completely autonomous anticipatory mobile computing system remains an interesting research question.

\subsubsection{\red{Filtering context sensed data for practical anticipatory computing}}
With an ever-increasing number of sensing modalities available on a phone, an anticipatory mobile system faces the problem of selecting useful features from a myriad of extracted sensor data. Psychological experiments show that the human mind is well versed in filtering out signals that are irrelevant for the task at hand \cite{Simons1999}. Machine learning tools often implicitly filter out unimportant signals, for example, a regression model weights factors according to their influence on the target variable. Sensing and processing sensor data, as well as machine learning modelling, require substantial resources, and explicit filtering, such as the one performed by human intuition, could improve the performance of artificial anticipatory systems. 

\subsubsection{\red{Defining the scope of anticipation}}
The amount of sensory input is not the only limitation that anticipatory systems should impose onto themselves. Anticipatory systems should be aware of the scope in which they operate and limit their liability to a specific time horizon and events within that scope. The time horizon of anticipation determines how far into future the anticipation goes. Setting the appropriate horizon is of great practical importance. A decision on whether to bring an umbrella or not is useless if made once we are already on the way to work. A decision to bring an umbrella on a day which is a year from now is likely to be inaccurate. Obviously, there is an inherent tradeoff between~\textit{accuracy} and \textit{curiosity} that anticipatory systems have to deal with. Highly proactive behaviour is useful only if the underlying predictions are correct. A sweet spot that determines how far in the future a decision should be made depends on the accuracy of the model, but also on the application for which predictions are made. 


\subsection{\red{Designing }Advanced Applications}

As the area matures, and the predictions that mobile devices afford become more reliable, we expect closer integration with fields that can directly benefit from anticipatory mobile applications. 
The most likely synergies can be envisioned in the areas of psychology and healthcare.
Predictions of future context can help with psychological therapy design and delivery~\cite{Lathia2013,Pejovic14:anticipatory}. On the other hand, behavioural theory can potentially be integrated with machine learning algorithms for more robust and reliable predictions. Information and communication technologies are key enablers of \textit{smart cities} -- efficient, sustainable urban environments. Rich computing and sensing capabilities together with smartphones' geographically limitless inter-connectivity allow both broad, society-wide, as well as individual context inference. In future smart cities, anticipatory mobile computing could manage crowds and traffic, help with environmental monitoring and protection, and be used as a basis for public safety applications.

Anticipatory mobile computing enables paradigm-changing opportunities for adaptive systems, as introduced in Section~\ref{sec:mobility_prediction}. It is possible to envision a network that adapts its connectivity to accommodate predicted usage surges. But even beyond adaptation to predicted physical context, systems can adapt to predicted inner state of the user. Psychological computing was proposed by Bao et al. as a class of computing systems that sense user's inner context and utilise it on the core system level~\cite{Bao2013}. Anticipatory computing systems based on predictions of users' internal state fulfil this definition of psychological computing. For example, we can also envision flexible content delivery systems that cache digital content according to predicted users' interests and emotional states.

\section{Summary}
\label{sec:summary}

In this article we discussed the nascent field of anticipatory mobile computing, and surveyed the key concepts on which it is founded.
Context inferences made through mobile sensing serve as a basis for predictive models on which anticipatory computing is based. In addition, anticipatory decisions are often delivered by the phone and executed by the user. Context-sensitive delivery, enabled by mobile sensing, is crucial for efficient actioning. In this survey we paid special attention to sensing, modelling and prediction of high-level context concepts. On the anticipatory computing side, our aim was to present reinforcement learning and latent learning as suitable solutions for the least intrusive, and efficient anticipatory mobile computing implementation. 

We note that the foundations of anticipatory mobile computing -- mobile sensing and anticipatory computing -- are continuously changing. Indeed, merely seven years have passed since Apple released its iPhone. In each subsequent generation smartphones have been equipped with a larger number of sensors and more powerful computational resources that allow to execute more powerful sensor data processing algorithms. 
Similarly, anticipatory computing is an active ever-changing research area driven by recent advances in diverse fields from robotics to psychology. Therefore, in this survey we concentrated on identifying state-of-the-art practical endeavours and promising research trends.  In addition, our goal was to extract best implementation practices and consolidate disjoint efforts from mobile sensing and anticipatory computing research communities. 

Surge of related research activity together with a number of recently released predictive applications, such as Google Now, Microsoft Cortana, Apple Siri, Yahoo Aviate and MindMeld, stand as evidence of the rising importance of anticipatory mobile computing. The smartphone is already capable of both society-wide and individual context inference and prediction. Once we merge phone's predictions with advance intelligence capable of steering the future through interaction with the user, a whole new set of ground-breaking applications might be possible.
It is our hope that the path laid out in this article represents a valuable guideline for further efforts in this fascinating emerging area.

\section*{Acknowledgements}
The authors thank Mihai Nadin from UT Dallas for a thoughtful discussion about anticipatory computing.

\bibliographystyle{ACM-Reference-Format-Journals}
\bibliography{references}
%
%
%

\end{document}